\documentclass[lettersize,journal]{IEEEtran}
\IEEEoverridecommandlockouts
\usepackage{cite}
\usepackage{amsmath,amssymb,amsfonts}
\usepackage{algorithmic}
\usepackage{graphicx,subfigure}
\usepackage{makecell}
\usepackage{textcomp}
\usepackage{tikz}
\usepackage{xcolor}
\usepackage{listings}
\usepackage{multirow}
\usepackage{titlesec}
\usepackage{orcidlink}

\titlespacing{\section}{0pt}{\parskip}{-\parskip}
\titlespacing{\subsection}{0pt}{\parskip}{-\parskip}
\titlespacing{\subsubsection}{0pt}{\parskip}{-\parskip}

\usepackage[all]{background}
\usepackage{stackengine}
\setstackEOL{\\}
\setstackgap{L}{\normalbaselineskip}
\SetBgContents{\color{gray}{\tiny \Longstack{PREPRINT - To Appear at IEEE ESL 2023}}}% Set contents
\SetBgPosition{4.5cm,1cm}% Select location
\SetBgOpacity{1.0}% Select opacity
\SetBgAngle{0}% Select rotation of logo
\SetBgScale{1.8}% Select scale factor of logo
\usepackage{enumitem}
\setlist[itemize]{align=parleft,left=0pt..1em}
\colorlet{punct}{red!60!black}
\definecolor{background}{HTML}{EEEEEE}
\definecolor{delim}{RGB}{20,105,176}
\colorlet{numb}{magenta!60!black}

\newcommand*\circled[1]{\tikz[baseline=(char.base)]{
            \node[shape=circle,fill,inner sep=1pt] (char) {\textcolor{white}{#1}};}}

\lstdefinelanguage{json}{
    basicstyle=\ttfamily\scriptsize,
    numberstyle=\tiny,
    numbersep=8pt,
    showstringspaces=false,
    breaklines=true,
    frame=lines,
    backgroundcolor=\color{background},
    literate=
     *{0}{{{\color{numb}0}}}{1}
      {1}{{{\color{numb}1}}}{1}
      {2}{{{\color{numb}2}}}{1}
      {3}{{{\color{numb}3}}}{1}
      {4}{{{\color{numb}4}}}{1}
      {5}{{{\color{numb}5}}}{1}
      {6}{{{\color{numb}6}}}{1}
      {7}{{{\color{numb}7}}}{1}
      {8}{{{\color{numb}8}}}{1}
      {9}{{{\color{numb}9}}}{1}
      {:}{{{\color{punct}{:}}}}{1}
      {,}{{{\color{punct}{,}}}}{1}
      {\{}{{{\color{delim}{\{}}}}{1}
      {\}}{{{\color{delim}{\}}}}}{1}
      {[}{{{\color{delim}{[}}}}{1}
      {]}{{{\color{delim}{]}}}}{1},
}

\def\BibTeX{{\rm B\kern-.05em{\sc i\kern-.025em b}\kern-.08em
    T\kern-.1667em\lower.7ex\hbox{E}\kern-.125emX}}
    
\begin{document}

\bstctlcite{IEEEexample:BSTcontrol}

%\title{Breakdown of Energy Consumption for Stateful Logic-in-Memory\vspace{-1.5cm}}
\title{Should We Even Optimize for Execution Energy? Rethinking Mapping for MAGIC Design Style}

\author{Simranjeet Singh\orcidlink{0000-0002-8297-1470}, Chandan Kumar Jha\orcidlink{0000-0002-7237-5878}, Ankit Bende\orcidlink{0009-0008-6434-7667}, Phrangboklang Lyngton Thangkhiew\orcidlink{0000-0001-8109-1458}, \\ Vikas Rana\orcidlink{0000-0001-5432-0286}, Sachin Patkar, Rolf Drechsler\orcidlink{0000-0002-9872-1740}, and Farhad Merchant\orcidlink{0000-0002-3708-5621} 
\vspace{-0.8cm}  

\thanks{This work was supported in part by the BMBF, Germany, in the project NEUROTEC II under Project 16ME0398K, Project 16ME0399, DFG within the Project PLiM (DR 287/35-1, DR 287/35-2) and through Dr. Suhas Pai Donation Fund at IIT~Bombay.}

\thanks{Simranjeet Singh and Sachin Patkar are with the IIT Bombay, Mumbai 400076, India, and Simranjeet Singh is also with Forschungszentrum Jülich
GmbH, PGI, Jülich 52425, Germany (e-mails: \{simranjeet, patkar\}@ee.iitb.ac.in).}
\thanks{Ankit Bende and Vikas Rana are with Forschungszentrum Jülich
GmbH, PGI, Jülich 52425, Germany  (e-mails: \{a.bende,v.rana\}@fz-juelich.de).} 
\thanks{Chandan Kumar Jha and Rolf Drechsler are with the Institute of Computer
Science, University of Bremen, Bremen 28359, Germany. Rolf Drechsler is also with the Department of Cyber-Physical Systems, DFKI GmbH, Bremen 28359, Germany (e-mail: \{chajha, drechsler\}@uni-bremen.de).}
\thanks{Phrangboklang Lyngton Thangkhiew is with the IIIT Guwahati, Assam 781015, India (e-mail: phrangboklang.thangkhiew@iiitg.ac.in).}
\thanks{Farhad Merchant is with Newcastle University, Newcastle upon Tyne NE17RU, United Kingdom (e-mail: farhad.merchant@newcastle.ac.uk). }}

\markboth{IEEE EMBEDDED SYSTEMS LETTER,~Vol.~x, No.~x, September~2023}%
{Shell \MakeLowercase{\textit{et al.}}: A Sample Article Using IEEEtran.cls for IEEE Journals}

\IEEEpubid{0000-0000~\copyright~2023 IEEE.}

% \author{\IEEEauthorblockN{Simranjeet Singh\IEEEauthorrefmark{1}\IEEEauthorrefmark{5}, Chandan Kumar Jha\IEEEauthorrefmark{3}, Ankit Bende\IEEEauthorrefmark{5}. Vikas Rana\IEEEauthorrefmark{5}, \\ Sachin Patkar\IEEEauthorrefmark{1}, Rolf Drechsler\IEEEauthorrefmark{3}\IEEEauthorrefmark{4},
% Farhad Merchant\IEEEauthorrefmark{2} \IEEEauthorblockA{\IEEEauthorrefmark{1}Indian Institute of Technology Bombay, India, \IEEEauthorrefmark{3}University of Bremen, Germany, \\ \IEEEauthorrefmark{5}Forschungszentrum Jülich GmbH, Germany, \IEEEauthorrefmark{4}DFKI GmbH, Germany,  \IEEEauthorrefmark{2}Newcastle University, UK}}
% \{simranjeet, patkar\}@ee.iitb.ac.in, \{si.singh, a.bende, v.rana\}@fz-juelich.de, \\ \{chajha, drechsler\}@uni-bremen.de, farhad.merchant@newcastle.ac.uk \vspace{-0.7cm}}

\maketitle
% \begingroup\renewcommand\thefootnote{\textsection}
% \footnotetext{Equal contribution}
% \endgroup
\begin{abstract}

% To the best of our knowledge, no framework allows the generation of a spice netlist from the hardware description language for IMC using memristors. The current state of the works is limited to only generating the mapping and the micro-operations that need to be performed. In this work, we, for the first time, propose a framework that generates the netlist and the testbench required for evaluating the memristor crossbar. 
% Logic-in Memory using memristors has become popular in alleviating the von Neumann bottleneck in conventional computing. However, Understanding the energy breakdown while designing the logic operation inside the memory is critical to analyze the importance of such implementation. The state-of-the-art energy estimation techniques for the mapping obtained using MAGIC-based designs are based on coarse-grain techniques. We show that the assumptions used in these coarse-grain techniques vastly underestimate the energy consumption of the MAGIC operations performed using the memristor crossbar. To alleviate this concern, we break down the energy consumed during logic-in memory, which takes the mapping obtained from the SIMPLER tool to perform accurate energy estimation using SPICE simulations. 

Memristor-based logic-in-memory (LiM) has become popular as a means to overcome the von Neumann bottleneck in traditional data-intensive computing. Recently, the memristor-aided logic (MAGIC) design style has gained immense traction for LiM due to its simplicity.  However, understanding the energy distribution during the design of logic operations within the memristive memory is crucial in assessing such an implementation's significance. The current energy estimation methods rely on coarse-grained techniques, which underestimate the energy consumption of MAGIC-styled operations performed on a memristor crossbar. To address this issue, we analyze the energy breakdown in MAGIC operations and propose a solution that utilizes mapping from the SIMPLER MAGIC tool to achieve accurate energy estimation through SPICE simulations. In contrast to existing research that primarily focuses on optimizing execution energy, our findings reveal that the memristor's initialization energy in the MAGIC design style is, on average, 68$\times$ higher. We demonstrate that this initialization energy significantly dominates the overall energy consumption. By highlighting this aspect, we aim to redirect the attention of designers towards developing algorithms and strategies that prioritize optimizations in initializations rather than execution for more effective energy savings.

% We highlight that contrary to existing works that focus on optimizing execution energy, initialization energy on average is 70$\times$ higher. We show that initialization energy in the MAGIC design style significantly dominates the overall energy consumption. We believe this work will help designers to refocus toward implementing algorithms to prioritize optimizations in initializations rather than execution for better energy savings. 

%This raises the need to optimize logic designs with fewer initialization steps instead of logic gates.
%will add the number after generating results

\end{abstract}

% \begin{IEEEkeywords}
% Memristors, Logic-in Memory, MAGIC
% \end{IEEEkeywords}



\maketitle
% \IEEEkeywords{PUF, TRNG, RRAM, Memristors, Hardware Security}
\section{Introduction}
% In-Memory Computing (IMC) using memristors has become popular in alleviating the von Neumann bottleneck in conventional Computing. Memristors have two states, a high resistive state (HRS) and a low resistive state (LRS), which are further mapped to boolean logic '0' and logic '1', respectively. These resistive states are used to design the Boolean logic gates inside the memory. Several approaches have been proposed in the literature to implement the logic gates using memristors such as MAGIC~\cite{KBL+:2014}, IMPLY\cite{BSK+:2010}, FELIX~\cite{gupta2018felix}, and Majority logic~\cite{DDH+:2023}. MAGIC is a widely adopted technique, as it outperforms other IMC techniques in terms of energy and latency. MAGIC design style proposes the implementation of NOR and NOT gates using a memristor crossbar. Figure 1 shows the MAGIC NOR and NOT implementation using memristors. The output of NOR and NOT can be stored in a dedicated output memristor without any special arrangement in the crossbar, which makes this technique suitable for digital IMC. 

% \textcolor{red}{PLEASE SHORTEN THE INTRODUCTION}
\IEEEPARstart{C}{omputing-in-memory} is one way of reducing the impact of the von Neumann bottleneck. As a result, digital logic-in-memory (LiM) has gained significant momentum recently. Memristive memories are seen as a viable candidate for LiM. Memristors possess two distinctive states: the high resistive state (HRS) and the low resistive state (LRS), which are then mapped to boolean logic '0' and logic '1', respectively. These resistive states are the foundation for designing Boolean logic gates within the memory. To model the memristive behavior in SPICE, various model has been proposed in the literature. The VTEAM model~\cite{KRE+:2015} is derived from the derivative of the internal state variables. This model can accurately represent the switching characteristics observed in memristor devices, making it a suitable choice for this study.

%which is a widely accepted approach in the literature. 
% VTEAM model~\cite{KRE+:2015} is one of the models that can characterize various memristive technology and is also used for this work.

Numerous approaches have been proposed in the literature to implement logic gates using memristors, including techniques such as memristor-aided logic (MAGIC)~\cite{KBL+:2014}, memristor-based material implication (IMPLY)~\cite{BSK+:2010}, fast and energy-efficient logic in memory (FELIX)~\cite{gupta2018felix}, and majority logic~\cite{DDH+:2023}.
MAGIC has emerged as a widely adopted technique among these approaches due to its superior energy efficiency and latency performance~\cite{EHA+:2021}.
% \textcolor{red}{IS THIS CORRECT?}. 

\begin{figure}[!t]
    \centering
    \subfigure[]{\includegraphics[width=0.5\linewidth]{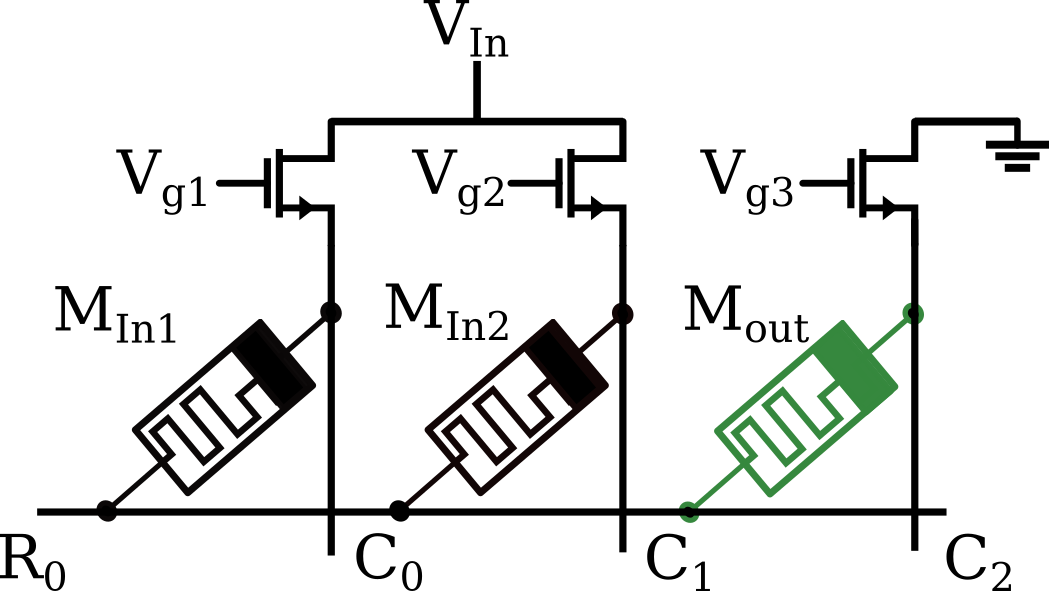}}
    \subfigure[]{\includegraphics[width=0.35\linewidth]{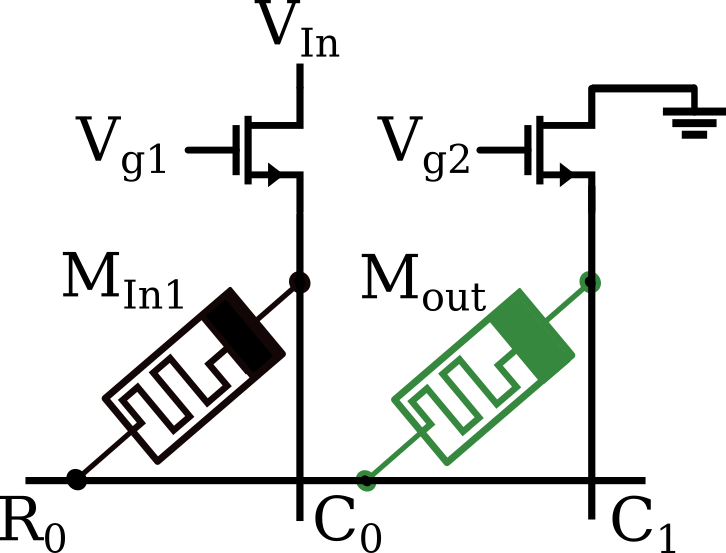}}
    \caption{MAGIC design style based (a) NOR gate, and (b) NOT gate}
    \vspace{-0.6cm}
    \label{nornot}
\end{figure}

MAGIC is a stateful logic technique to implement logic operations using memristive devices, where inputs and outputs of logic operations are stored in the resistive states of memristors. Fig.~\ref{nornot} visually represents the implementation of MAGIC NOR (1a) and NOT (1b) gates using memristors. An output memristor, $M_{out}$, is initialized to LRS, and it changes the state from LRS to HRS based on the input state stored in $M{in1}$ and $M{in2}$. NOR and NOT operations use three (two input, one output) and two memristors (one input, one output) in a single operation, respectively. The output of these logic gates can be conveniently stored in a dedicated output memristor without requiring any specific arrangement in a crossbar, making this technique particularly suitable for digital LiM applications. The state-of-the-art tool in this field, known as the SIMPLER MAGIC~\cite{RRA:2020}, is designed explicitly for synthesizing the MAGIC NOT and NOR operations into a single-row memristor crossbar. Henceforth, we refer to SIMPLER MAGIC as SIMPLER.
\IEEEpubidadjcol

The utilization of the MAGIC design style and its mapping onto the crossbar has been suggested for creating an in-memory general-purpose processing unit, mMPU~\cite{EHA+:2021}. As this design style is rapidly gaining popularity in mainstream computing, assessing the amount of energy consumed by this technique is crucial. The current methods for calculating energy consumption involve multiplying the average energy used during an operation by the number of such operations in an application, which is a highly coarse-grained approach to determine the energy consumed by the MAGIC design style~\cite{thangkhiew2018efficient}. Surprisingly, despite its popularity, this methodology falls short of providing accurate estimates of the energy dissipated by an application since it does not account for the energy consumed during initialization, reading, and loading input patterns.

%jump to paper contribution
% This paper proposes spiceMAGIC, an automated spice-level simulation and accurate energy estimation framework for the MAGIC design style. The proposed framework automatically generates a SPICE-level netlist and testbench voltages for a given application. Furthermore, it provides fine-grained energy numbers by calculating the energy consumed by each device in the crossbar, irrespective of its contribution to the operation.

This letter shows a novel and accurate \emph{fine-grained} methodology for energy calculation of LiM, where we break down the energy consumption for the MAGIC design style. We generate a SPICE-level netlist and testbench voltages for a given application to find the accurate energy breakdown. Furthermore, it provides fine-grained energy numbers by calculating the energy consumed by each device in the crossbar, irrespective of its contribution to the operation.

% \begin{itemize}
%     % \item We propose the spiceMAGIC framework, which takes SIMPLER tool mapping (.JSON) as an input and automatically generates the SPICE-level netlist. It allows the facility to change the device model and variations if necessary. Moreover, spiceMAGIC generates the voltage sequence for a given mapping for benchmarking.
%     \item Firstly, a fine-grained energy breakdown for MAGIC design style implementation in a crossbar \textcolor{red}{THIS IS NOT WHAT YOU ARE IMPLEMENTING IN A CROSSBAR}. Each device's consumption in the crossbar is measured for accurate energy consumption, irrespective of its contribution to the operation.
%     \item Lastly, we generate the SPICE-level netlist for widely used benchmarks (ISCAS'85) and compare the energy consumption with current methods.
%     \textcolor{red}{SPLIT these two bullet points into three contributions}
% \end{itemize}

% \textcolor{red}{Check these bullet points.}

\begin{itemize}
    \item Firstly, a fine-grained methodology for energy calculation for MAGIC design style. (Section~\ref{methodology})
    \item A detailed evaluation for energy consumption where each device's consumption is measured in a crossbar, irrespective of its contribution to the operation. (Section~\ref{experiments})
    \item Lastly, The discussion on approx. 70x average energy gap for ISCAS'85 benchmarks when compared to the state-of-the-art energy calculation methods. (Section~\ref{discussion} and~\ref{conclusion})
\end{itemize}
% --In-Memory computing, emphasis on MAGIC \\
% -- Energy techniques \\
% -- Emphasis on the energy consumption due to isolation voltages \\
% -- Motivation for our proposed work \\

% \textcolor{red}{I think you can get rid of the paper organization. You can mention the section numbers in a bracket next to the contribution bullets.}The rest of the paper is organized as follows. We elaborate on the methodology in Section~\ref{methodology}. In Section~\ref{experiments}, we discuss the benchmark circuit preparation and energy breakdown results and compare the results with state-of-the-art methods. We conclude the paper in Section~\ref{conclusion}.

% \section{Background and Related Work}
% \label{backnrw}
% \input{Sections/rw.tex}
\begin{figure}[t!]
    \centering
    \includegraphics[width=0.9\linewidth]{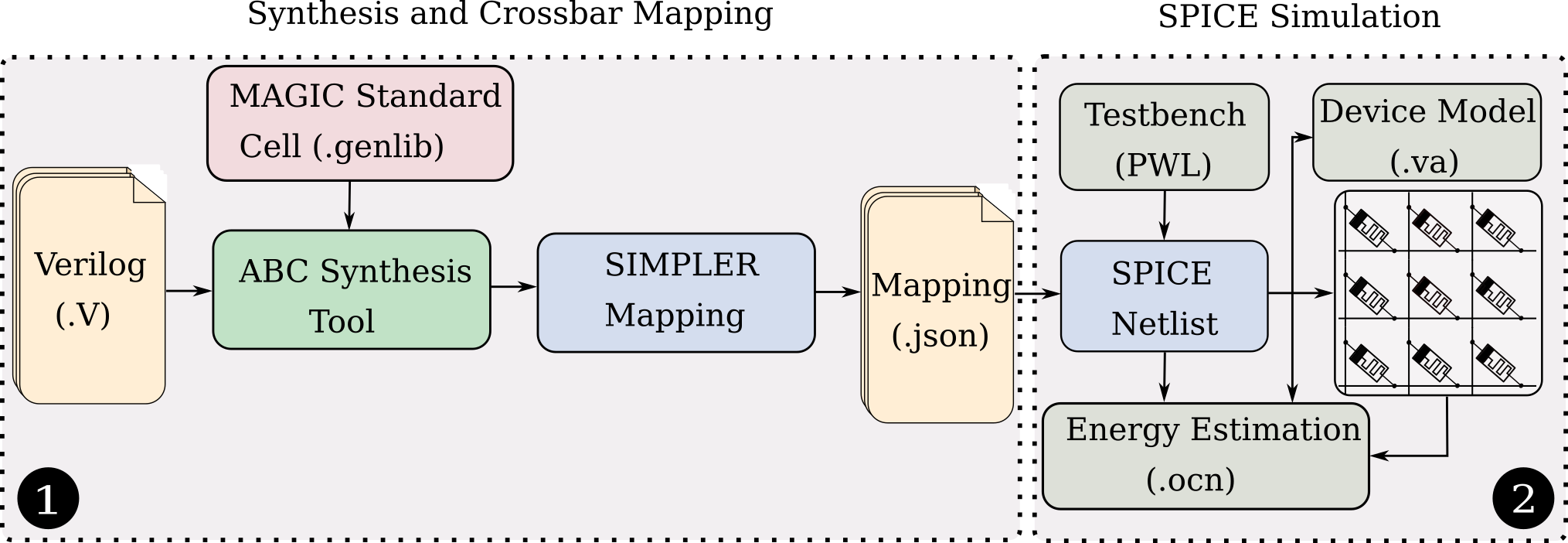}
    \caption{Methodology for generating a SPICE-level netlist and estimating energy for any given workload.\vspace{-1.2cm}}
    \label{fig:method}
\end{figure}

\section{Methodology}
\label{methodology}

% This section presents the SPICE MAGIC methodology to obtain accurate energy numbers by mapping the Verilog design to the MAGIC design style at the SPICE level. The SPICE MAGIC methodology is shown in Fig~\ref{fig:method}. It is an automated tool to generate the SPICE-level netlist for MAGIC NOR and NOT gate. The automated process comprises three-step (1) SIMPLER-tool mapping, (2) automated SPICE level netlist creation, and (3) automated testbench and energy estimations. In the following section, we discuss each step in detail.
% \textcolor{red}{Never start or say this section, that section.}This section discusses the netlist generation, testbench generation, and energy calculation techniques. The overall methodology is shown in Fig~\ref{fig:method}. We will now discuss the flow in detail.

Overall energy calculation methodology comprises three steps (see Fig.\ref{fig:method}): i) synthesis and crossbar mapping, ii) SPICE netlist generation, and iii) testbench and energy and estimation.  

\subsection{Synthesis and Crossbar Mapping}
% \textcolor{red}{This section heading should be Netlist Generation according to the preamble. What you are promising and what you are supplying are orthogonal.}
% The leading-edge tool in this field, known as the SIMPLER tool~\cite{RRA:2020}, is specifically designed for synthesizing the MAGIC NOT and NOR operations into a single-row memristor crossbar.
We begin the process by synthesizing the Verilog design using the ABC synthesis tool~\cite{mishchenko2007abc}. The ABC tool allows to synthesis of any arbitrary logic function into NOR and NOT logic gates as shown in Fig.~\ref{fig:method}~\circled{1}. Subsequently, the NOR/NOT netlist serves as input for the SIMPLER mapping tool, generating an optimal mapping for MAGIC design style gates. Moreover, the SIMPLER mapping tool performs a sequential mapping of the MAGIC NOT and NOR operations, which require the utilization of three and two memristors on the crossbar, respectively. Additionally, the SIMPLER tool generates the necessary information, such as the required number of cycles, input/output memristors, and other relevant details specific to the application or benchmark, which are then stored in a .json file. Listing~\ref{lst:half_adder} shows the .json of half adder mapping on five memristors connected in a single row. Moving forward, the .json file has been used to generate the SPICE-level netlist and test vectors. 
\subsection{SPICE Netlist Generation}
% Numerous models have been proposed in the literature to characterize the memristive devices for SPICE simulation. VTEAM model~\cite{KRE+:2015} is one of the models that can characterize various memristive technology and are used for this work.
% To generate the SPICE-level netlist, the crossbar mapping (.json) file is used to extract critical information such as input and output data devices and execution sequences of NOR/NOT operations. Based on the number of memristors required to implement the given set of applications, The crossbar netlist is designed by taking the VTEAM model as a memristors devices. All the operations are mapped to a single row of the crossbar, which has 'n' memristors connected to columns. The row of the crossbar is connected to the ground via a controllable switch. During the writing process, the row is connected to GND otherwise, it remains floating as a requirement of MAGIC execution. Considering all the parameters, a spectre-compatible netlist (.scs) in a crossbar structure is designed. Further, the crossbar SPICE netlist is packed into a crossbar symbol with dedicated input and output for benchmarking.

As shown in Fig.~\ref{fig:method}~\circled{2}, SPICE simulation takes input from the synthesis and mapping block to generate the SPICE-level netlist. In the SPICE simulation block, the test bench containing voltage files and energy estimation scripts is also created for accurate energy estimation. The critical information is extracted from the crossbar mapping (.json) file. This information includes the input and output data devices and the execution sequences of NOR/NOT operations. Based on the required number of memristors for the given set of applications, the crossbar netlist is designed using the VTEAM model as memristor devices. All operations are mapped to a single row of the crossbar, where `n' memristors are connected to columns. A controllable switch connects the row of the crossbar to the ground during the writing process while it remains floating as a requirement for MAGIC execution. Considering all the relevant parameters, a spectre-compatible netlist (.scs) is designed in the structure of a crossbar. Subsequently, the crossbar SPICE netlist is encapsulated within a crossbar symbol with dedicated input and output for benchmarking purposes.
\begin{lstlisting}[language=json,caption={Mapping of a half adder onto five memristors in a row.},captionpos=b,label={lst:half_adder},frame=single,linewidth=0.98\linewidth]
"Row size": 5,
"Number of Gates": 5,
"Inputs": "{A(0),B(1)}",
"Outputs": "{S(4),Cout(2)}",
"Reuse cycles": 1,
"Execution sequence": {
"T0": "Init{'D(2)','D(3)','D(4)'}",
"T1": "n5_(4)=inv1{A(0)}",
"T2": "n6_(3)=inv1{B(1)}",
"T3": "Cout(2)=nor2{n6_(3),n5_(4)}",
"T4": "Init{n5_(4),n6_(3)}",
"T5": "n8_(3)=nor2{B(1),A(0)}",
"T6": "S(4)=nor2{n8_(3),Cout(2)}"}

\end{lstlisting}

% It also considers the device model (verilog.a model or path of std cell) to map the operation on the device model. It also provides the features to consider the variations of the devices during execution sequences, such as device-to-device variation and cycle-to-cycle.

% To generate the automated SPICE-level netlist, spiceMAGIC considers the output of the SIMPLER tool (.json) as an input and extracts the critical information such as input and output data devices and execution sequences of NOR/NOT operations. It also considers the device model (verilog.a model or path of std cell) to map the operation on the device model. It also provides the features to consider the variations of the devices during execution sequences, such as device-to-device variation and cycle-to-cycle. 

% As the SIMPLER tool maps the design to a single row, the generated SPICE-level netlist also contains a single row with 'n' numbers of columns. Further, the crossbar SPICE netlist is packed into a crossbar symbol with dedicated input and output for benchmarking.

\begin{figure}
    \centering
    \includegraphics[width=0.85\linewidth]{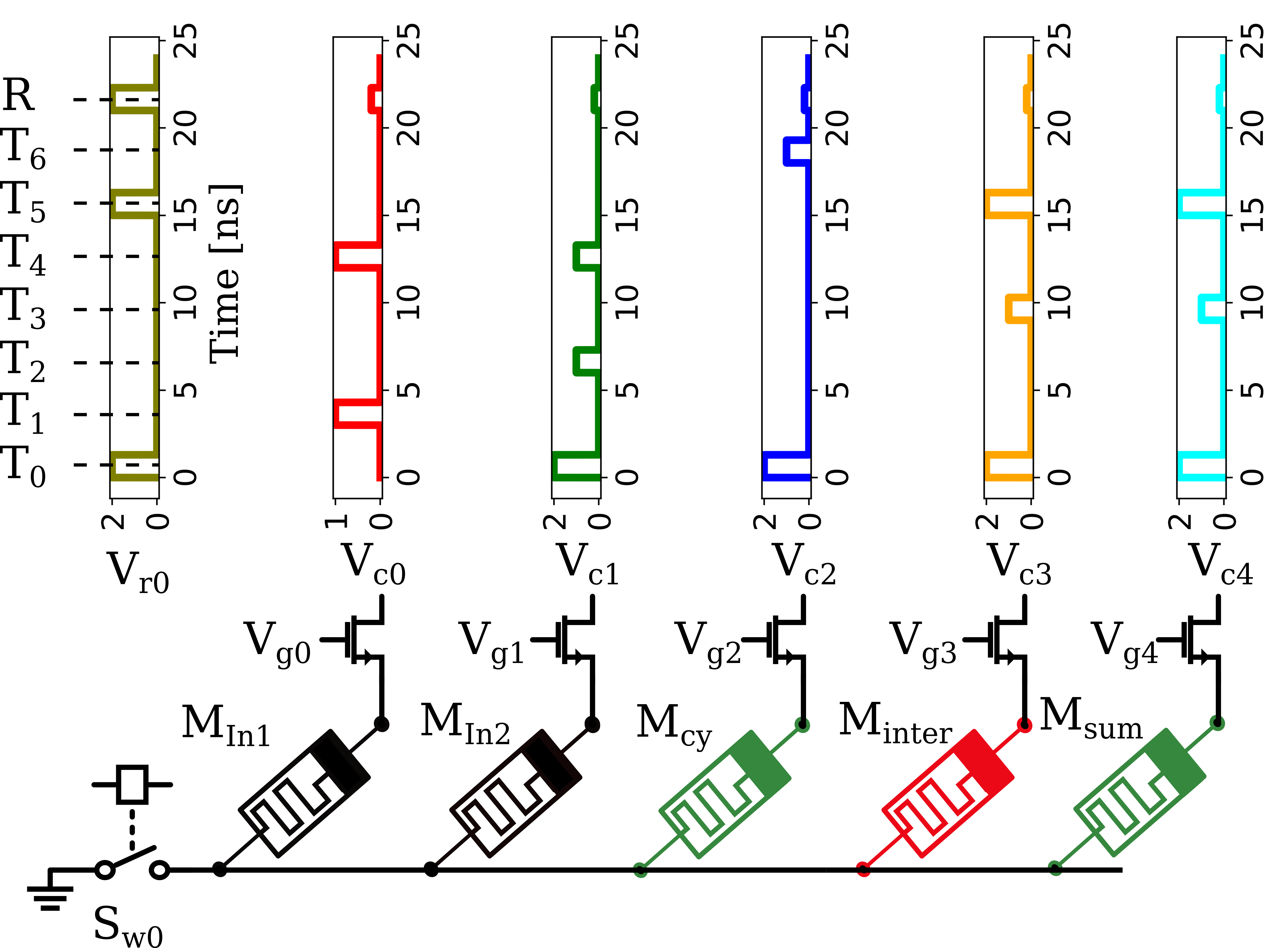}
    \caption{Implementation of a half adder using five memristors arranged in a row, based on the mappings provided in Listing~\ref{lst:half_adder}. The waveform configuration involves a total of 7 execution cycles ($T_0$-$T_6$), including an initialization cycle ($T_0$), a reused cycle ($T_4$), and a final read cycle (`R') to observe the output states of each device.\vspace{-5mm}}
    \label{fig:waveform}
\end{figure}

\subsection{Testbench and Energy Estimation}
To enable the operations in digital in-memory computing, input voltages play a crucial role in configuring the functionality. Executing any arbitrary digital logic on the memristor crossbar becomes possible by applying different voltage combinations with specific values to the rows and columns. In digital LiM using memristors, there are five distinct voltage levels required:
\begin{itemize}
    \item \textbf{\textit{Input Voltage:}} This voltage determines the resistances of the memristors used as inputs. The input voltage is mapped to 2.0V for storing `1' (LRS) and 0.0V for HRS (by default, the memristors are in HRS).

    \item \textbf{\textit{Read Voltage:}} Applied to read the state of the output memristors after all operations have been performed. The read voltage is relatively low, set at 0.2V in this case, compared to the SET and RESET voltages. The read voltage is chosen to ensure reliable read operations; however, there is potential for reducing energy consumption by reducing the read voltage.

    \item \textbf{\textit{Initialization Voltage:}} The intermediate output memristors, which store intermediate results, need to be initialized to LRS before accurate operation. This voltage configures the intermediate output memristors to LRS and is mapped to 2.0V for correct operation.

    \item \textbf{\textit{Gate Voltage:}} During the execution cycle, only a few devices are chosen for operation, while others in the crossbar must be isolated. The gate voltage is used to select the devices for operation. The gate voltage of the selected devices is set to 2.0V (ON), while the others are kept at 0V (OFF). This voltage ensures that the state of these memristors remains unaltered during computation.

    \item \textbf{\textit{Operation Voltage:}} During operation, a specific voltage is applied to the memristors with input values to execute MAGIC design style-based NOR and NOT operations. A voltage of 1.0V is applied to one memristor during the NOT operation and to two memristors during the NOR operation. The intermediate output memristor is connected to the ground.
\end{itemize}
\begin{table}[!b]
    \centering
    \caption{MAGIC NOR, NOT and Write Energy Estimation}
    \label{tab:nor_not_write_table}
    \begin{tabular}{|c|c|c|}
    \hline
  NOR, NOT\&  &  $V_{in}=1.0$@1.3 ns  & $V_{in}=1.0$@1.3 ns \\
    Write Operation&  DC source  & rise and fall time (1 ps)\\
  \hline
       \textbf{NOR}   & Energy (fJ) &  Energy (fJ)  \\
       \hline
        00 $\rightarrow$ 1  & 8.6 &  8.6 \\ 
        \hline
        01  $\rightarrow$ 0 & 89.53 &  87.96 \\ 
        \hline
        10  $\rightarrow$ 0 & 89.53 &  87.96 \\ 
        \hline
        11  $\rightarrow$ 0 & 32.48 & 30.48 \\ 
        \hline
        \textbf{Average} & \textbf{55.04} & \textbf{53.75} \\
        \hline
        \hline
        % NOT & & \\
        \textbf{NOT}   & Energy (fJ) &  Energy (fJ)  \\
       \hline
        0 $\rightarrow$ 1  & 4.31 &  4.32 \\ 
        \hline
        1  $\rightarrow$ 0& 90.31 &  88.74 \\ 
        \hline
        \textbf{Average} & \textbf{47.31} & \textbf{46.53} \\
        \hline
        \hline
       \textbf{Writing}  & Energy (fJ) @1.0 ns &  Energy (fJ)  \\
       \hline
        RESET $\rightarrow$ SET (2V) & 75.56 &  \textbf{1272.2} \\ 
        \hline
        SET $\rightarrow$ RESET (1V) & 17.66 &  \textbf{19.01} \\ 
        \hline

    \end{tabular}
    
\end{table}
The implementation of these voltage values involves the use of a piece-wise linear voltage source. The determination of specific voltage levels is based on a single execution cycle provided in the .json file. Fig.~\ref{fig:waveform} demonstrates the voltage waveform explicitly designed for implementing an in-memory half adder, as depicted in Listing~\ref{lst:half_adder}. The SPICE-level netlist, which incorporates the VTEAM memristor model (.va) and the corresponding voltage sources, is compatible with Cadence spectre simulation.

\emph{Energy Estimation: } To accurately calculate the energy consumption, we perform a simulation for the required duration and generate a waveform file. The energy is calculated regardless of the device's selection. The total energy is determined using $\rm \sum_{i=0}^{n} \int_{0}^{t} (V_i \times I_i) dt$, which sums up the product of voltage and current over simulation time.
% \begin{equation}
% \label{equ:eng}
%  % \resizebox{0.3\textwidth}{!}{
% \rm Energy =\sum_{i=0}^{n} \int_{0}^{t} (V_i \times I_i) dt
% % }
% \end{equation}
Here, `n' represents the number of memristors, and `t' corresponds to the simulation time. The simulation time depends on the pulse width and the total number of cycles necessary to complete the benchmark. The given equation captures the activity on each memristor irrespective of its use in the cycle, which includes initialization, execution, and read energy.  

This section provided a detailed discussion on netlist generation, testbench generation, and energy estimation techniques for a specific benchmark. Furthermore, the simulated design is evaluated using the described methodology to assess its energy efficiency, and the results are further compared with state-of-the-art methodologies.

% In spiceMAGIC methodology, we have generated the voltage values at every time step for all the memristors. These voltage values are implemented using a piece-wise linear (PWL) voltage source in Cadence. \\

% \textcolor{red}{Show how the voltage file looks like }
% -- How we are generating PWL file (testbench) \\
% -- Ocean script to calculate the energy \\
% \textcolor{red}{Show the ocean script template}
% -- energy of each cell in every cycle. \\
% \textcolor{red}{Show the current and voltage waveform. Also, show the integral equation used in cadence}

\section{Experimental Results}

\label{experiments}
%add experiments results
% In this section, we discuss the results obtained using the spiceMAGIC framework. 
% The current methods for calculating energy consumption involve multiplying the average energy used during an operation by the number of such operations in an application, which is a highly coarse-grained approach to determine the energy consumed by the MAGIC design style~\cite{thangkhiew2018efficient}. Moreover, in literature, energy is calculated by applying a DC source for execution and the reported energy had been calculated at the exact switching point. However, in real implementation, we need a pulse width with a fixed amplitude for operation. Table~\ref{tab:nor_table} shows the energy presented in literature and energy for our methodology. Table~\ref{tab:nor_table} shows that execution energy for NOT implementation is almost matches with the presented energy values. Similarly, Table~\ref{tab:not_table} shows the comparison of energy with literature and proposed methodology, which also matches with reported results in the literature.
% \begin{table}[!b]
%     \centering
%     \caption{Energy for WRITE operation }
%     \label{tab:write_table}
%     \begin{tabular}{|c|c|c|c|}
%     \hline
%    &  & 1ns  & 1.3ns \\
%    &  & DC source  & rise/fall time (1ps)\\
%   \hline
% %   \hline
%        Operation & Voltage (V) & Energy (fJ) &  Energy (fJ)  \\
%        \hline
%         RESET $\rightarrow$ SET  & 2 & 75.56 &  1272.2 \\ 
%         \hline
%         SET $\rightarrow$ RESET & 1 & 17.66 &  19.01 \\ 
%         \hline
%     \end{tabular}
    
% \end{table}
\begin{table*}[!t]
    \caption{Energy Consumption Results on ISCAS'85 Benchmarks}
    \label{tab:benchmark}
    \setlength{\tabcolsep}{4pt}
    \centering
    \begin{tabular}{|c|c|c|c|c|c|c|c|c|c|c|c|c|c|c|c|c|}
    \hline
          Circuit & PI/PO & Cycles & NOT & NOR& \makecell{Energy (pJ) } &\multicolumn{3}{c|}{ \textbf{P1:} Energy (pJ), Input; all 0} &\multicolumn{3}{c|}{\textbf{P2:} Energy (pJ), Input; all 1} & \multicolumn{3}{c|}{\textbf{P3:} Energy (pJ), Input; alt.}\\
         \cline{7-15}
        & & & & &Literature~\cite{jha2022,thangkhiew2018efficient} &  Init & Exe & Read &  Init  & Exe & Read &  Init & Exe & Read\\
         \hline
         c17 & 5/2 & 14 &7 &6 & 0.655&  1161 & 0.674 & 11.55 & 1164 & 0.90 & 11.55 & 1162 & 0.75 & 11.56\\
        c432 & 36/7 &250 &101 &148& 12.31 &1142 & 15.23& 9.73&1163 & 15.96& 9.95&1153 & 14.92& 10.04 \\
        c499 & 41/32 & 605 &213 &390& 29.6&1790 & 30.85&6.23&1803 & 25.53&7.50 &1788 & 28.45&6.73\\
          c880 & 60/26 & 505 &194 &310 & 24.85& 1819& 22.91 & 7.25& 1785 & 25.26 & 8.024& 1779 & 24.76 & 7.504\\
          c1908 & 33/25 & 571 &210 &359& 27.99 &1850 &26.33 &7.13& 1814 & 22.95 & 7.701& 1822 & 25.72 & 7.39 \\
          c3540 & 50/22 & 1397 &465 &928 & 68.18&2676 &37.92 &4.41&2474 &37.51 &5.41&2431 &38.22 &4.84 \\
        \hline
        % \multirow{11}{*}{\textbf{ISCAS'89}} & c27 & 7/4 & 12 &5 &7 & 0.5944&  127 & 126.9 & 126.39\\
        %  & c208 & 19/10 & 100 &45 &55 & 4.977&  142.77 & 144.25 & 143.315\\
        % & c344 & 24/26 & 156 &60 &96 & 7.691&  154.61 & 155.74 & 155.239\\
        % & c386 & 13/13 & 185 &62 &123 & 9.055&  157.205 & 157.205 & 159.92\\
        % & c420 & 35/18 & 198 &87 &111 & 9.84&  163.93 & 164.5 & 162.68\\
        % & c510 & 25/13 & 312 &115 &197 & 15.34&  187.115 & 182.18 & 184.409\\
        % & c820 & 23/25 & 367 &136 &231 & 18.05&  198.622 & 191.002 & 194.431\\
        % & c832 & 23/24 & 347 &120 &227 & 17.01&  195.60 & 187.473 & 191.10\\
        % & c838 & 67/34 & 389 &166 &223 & 19.29&  204.16 & 202.23 & 203.59\\
        % & c1488 & 14/25 & 722 &207 &514 & 35.04&  1133.08 & 1108.63 & 1118.88\\
        % & c1494 & 14/25 & 737 &208 &528 & 35.74&  1133.08 & 1116.11 & 1125.31\\
        % \hline
    \end{tabular}
    
\end{table*}
The state-of-the-art methods used to calculate energy consumption in MAGIC design style applications rely on a coarse-grained approach, multiplying the average energy per operation by the total number of operations in the application. As discussed in~\cite{thangkhiew2018efficient}, this approach does not accurately capture the energy consumed by the MAGIC design style. Additionally, existing literature often calculates energy by applying a DC source for execution and measuring energy at the exact switching point. However, a pulse is required for operation in real implementations. In Table~\ref{tab:nor_not_write_table}, we compare the energy values obtained from the literature with the energy values calculated using our methodology. The results show that the execution energy for both NOT and NOR implementations closely matches the reported energy values in the literature~\cite{jha2022}.

In the MAGIC design style, energy consumption is predominantly dominated by the writing phase. Table~\ref{tab:nor_not_write_table} demonstrates that the writing process consumes 16.8$\times$ more energy during SET operation than the reported energy values, considering a pulse width of 1.3 ns and rise/fall times of 1 ps. Our proposed methodology uses a pulse width of 1.3 ns for writing since it is also the minimum required pulse width for correct operations. We want to highlight that energy consumption shoots up significantly if a larger pulse width is used. Due to the memristor's low resistance, it draws a significant current. To simplify the design of peripherals, a single pulse width is utilized, determined based on the worst-case pulse requirement in design.

The overall energy results are shown in Table~\ref{tab:benchmark}. The first and second columns have the benchmark suite's names and their respective benchmark circuits, respectively. The PI/PO column gives the number of primary inputs and primary outputs. The `Cycles' column gives the number of cycles required to obtain the final result for the given benchmark circuit. The `NOR' and the `NOT' columns give the number of the NOR and NOT operations, respectively. The energy consumption using the current state-of-the-art method is shown in the next column. In the last three columns, we show the energy consumption of the various benchmark circuit using three different input patterns, P1, P2, and P3, respectively. The pattern P1 denotes all 0's at the input, the pattern P2 denotes all 1's at the input, and the pattern P3 denotes alternating 1's and 0's at the input. We now discuss the results in detail using some examples from the benchmarks.

Fig.~\ref{fig:eng} illustrates the energy breakdown for the c3540 benchmark, with a focus on c3540 due to its larger size, which requires more initialization compared to other benchmarks. We see that the c3540 benchmark circuit from ISCAS 85 (Table~\ref{tab:benchmark}) has 50 inputs and 22 outputs. The number of cycles required for the operation is 1397. The circuit consists of 465 NOT gates and 928 NOR gates. The current methods that only evaluate the energy of the operation give 68.18 pJ as the overall energy consumption. We obtained the energy consumption with SPICE netlist for P1, P2, and P3 to be {2676 pJ, 2474 pJ, and 2431 pJ, respectively. The difference in the energy consumption value is approx. 40$\times$ compared to the current state of the art. The difference in energy consumption compared to the state-of-the-art method is lower than the c17 benchmark circuit as the design is larger than the entire crossbar and reuses the crossbar to perform the operations. The difference in the energy consumption between P1, P2, and P3 highlights that the operation energy dominates the initialization energy, and depending upon the input pattern, different energy values will be obtained. This allows us to capture the energy consumption of the design depending on the input patterns. 
\begin{figure}
    \centering
    \includegraphics[width=0.95\linewidth]{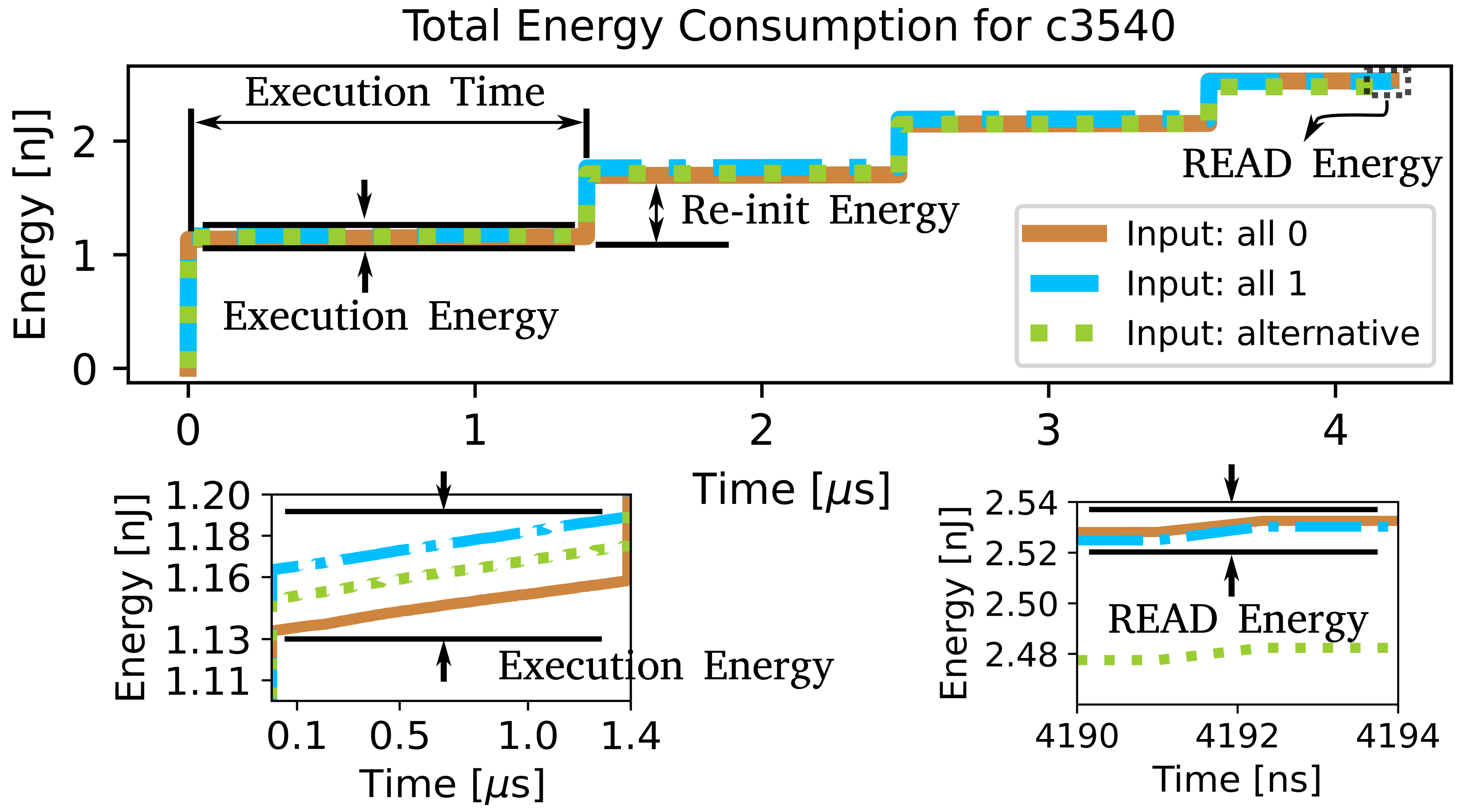}
    \caption{Energy breakdown and execution time analysis for the c3540 benchmark, highlighting different energy consumption components. The zoomed-in versions of the execution and read energy are highlighted in separate graphs.}
    \vspace{-0.5cm}
    \label{fig:eng}
\end{figure}

% \color[red]{change it}
The results for all the other benchmarks are shown in Table~\ref{tab:benchmark}. The energy values obtained using our methodology are very different than the ones obtained using the state-of-the-art energy calculation methodology. On average, the energy consumption of benchmark c432 to c3540 is 68$\times$ higher than the energy values presented in the literature.

\section{Discussion}
\label{discussion}
% -provide the insights, why the initialization energy is dominating and where the researcher wshould focus on to optimize the design \\
% - Also, we can discuss about the number of devices we can use in single row if we argue in not reusing the cycle.  \\
% - Even in SIMD implementation devices need to reinit for large design. 

% As per experimental results, the initialization energy dominates during logic implementation in the memory, which is normally considered one-time energy consumption.  However, for resource constraints implementation, initialization energy will dominate the energy. One can argue that it can be reduced by reducing the pulse width to the border of switching. Even in that case, the revitalization energy will dominate the overall energy. For example, If we consider we have to reinitialize one memristor for reuse. As per the current algorithm, we directly apply a SET pulse without reading the present state of the device to switch the device into LRS. If the selected device is already in LRS, the high current will flow from the device, leading to high energy consumption. One could add one extra cycle before the re-initialize cycle to read the device's current state. Only the devices which are in HRS can be switched back to LRS to reduce the energy. This comes with an extra cycle. In conclusion, the energy of any given benchmark depends on how many re-initialized cycles are in there, and even adding an extra read cycle depends on the number of devices in LRS before the re-initialization cycle.

Based on SPICE simulation results, the energy consumed during logic implementation in memory is primarily dominated by the initialization process, typically considered a one-time energy consumption. However, in resource-constrained implementations, the initialization energy becomes the dominant factor. It may be argued that reducing the pulse width towards the switching threshold can lower energy consumption. Nevertheless, even in such cases, the re-initialization energy will still dominate the overall energy consumption. We believe that this insight can be incorporated while developing efficient mapping algorithms, which is missing in the current works.

% For instance, when re-initializing a memristor for reuse, the current algorithm directly applies a SET pulse without checking the device's present state. A high current will flow if the selected device is already in the LRS, resulting in significant energy consumption. To mitigate this, an extra cycle can be added before the re-initialization cycle to read the device's current state.
 % Only devices in the HRS can then be switched back to LRS, reducing energy consumption. A high current will flow if the selected device is already in the LRS, resulting in significant energy consumption during re-initialization.

% However, this additional cycle 
% comes with its own READ energy cost. 

In conclusion, the initialization energy dominates the energy consumption of a given benchmark.
%, which needs to be considered. 
Secondly, the read energy needs to be considered. Contrary to prior works, we saw that execution energy is negligible as compared to initialization and read energy. Since the majority of prior works focus on reducing execution energy, we believe our work will pave the way for further research to optimize mapping strategies that prioritize initialization.
%and the presence of devices in LRS before the re-initialization cycle.

\section{Conclusion}
\label{conclusion}
% The utilization of memristors in Logic-in Memory has gained popularity as a means to overcome the von Neumann bottleneck found in traditional computing. However, it is crucial to understand the energy distribution during the design of logic operations within the memory to evaluate the significance of such an implementation. The current cutting-edge methods for estimating energy, which relies on coarse-grain techniques, are employed for mapping obtained through MAGIC-based designs. Our research reveals that the assumptions made in these coarse-grain techniques significantly underestimate the energy consumption associated with the MAGIC operations conducted using the memristor crossbar. To address this issue, we analyze the energy breakdown during Logic-in Memory and propose a solution that utilizes the mapping obtained from the SIMPLER tool to achieve accurate energy estimation through SPICE simulations.

This work introduces a methodology for accurately calculating the fine-grained energy consumption of logic operations in the MAGIC design style. Through SPICE simulation, we demonstrate that initialization energy significantly dominates the overall energy in MAGIC design style implementation, accounting on average for 68$\times$ more energy consumption as compared to presented state-of-the-art energy values. These findings emphasize the need for researchers to prioritize the optimization of initialization rather than execution. In future studies, we aim to develop mapping techniques further to optimize energy consumption in the MAGIC design style. 

% \begin{acks}
% here write the ack
% \end{acks}

\bibliographystyle{IEEEtran}
\bibliography{Bib}
\end{document}